\documentclass[aps,prl,twocolumn,superscriptaddress,amssymb]{revtex4}
\usepackage{graphicx}
\usepackage{amsmath}
\usepackage{amsfonts}
\usepackage{epsfig}
\usepackage{bm}
\usepackage{color}

\begin{document}

\title{Holon Wigner Crystal in a Lightly Doped Kagome Quantum Spin Liquid}
\author{Hong-Chen Jiang}
\email{hcjiang@stanford.edu}
\affiliation{Stanford Institute for Materials and Energy Sciences, SLAC and Stanford University, Menlo Park, California 94025, USA}
\author{T. Devereaux}
\email{tpd@stanford.edu}
\affiliation{Stanford Institute for Materials and Energy Sciences, SLAC and Stanford University, Menlo Park, California 94025, USA}
\author{S. A. Kivelson}
\email{kivelson@stanford.edu}
\affiliation{Department of Physics, Stanford University, Stanford, California,  94305, USA}
\date{\today}

\begin{abstract}
We address the problem of a lightly doped spin-liquid through a large-scale density-matrix renormalization group (DMRG) study of the $t$-$J$ model on a Kagome lattice with a small but non-zero concentration, $\delta$, of doped holes. It is now widely accepted that the undoped ($\delta=0$) spin 1/2 Heisenberg antiferromagnet has a spin-liquid groundstate. Theoretical arguments have been presented that light doping of such a spin-liquid could give rise to a high temperature superconductor or an exotic topological Fermi liquid metal (FL$^\ast$).  Instead, we infer that the doped holes form an insulating charge-density wave state with one doped-hole per unit cell - {\it i.e.} a Wigner crystal (WC).  Spin correlations remain short-ranged, as in the spin-liquid parent state, from which we infer that the state is a crystal of spinless holons (WC$^\ast$), rather than of holes. Our results may be relevant to Kagome lattice  Herbertsmithite upon doping.
\end{abstract}
\maketitle

\textbf{Introduction:} %
Broad interest in quantum spin liquid phases (QSLs) was triggered by the notion that they can be viewed as  insulating phases with preexisting electron-pairs, such that upon light doping  they might automatically yield high temperature superconductivity.\cite{Anderson_RVB_1987,Kivelson_PRB_1987,Rokhsar_PRL_1988,Laughlin_PRL_1988,Wen_PRL_1996,Balents2010,Senthil_PRB_2005} 
It has also been proposed that a doped QSL might form an exotic topologically ordered Fermi liquid state (known as an $\rm FL^\ast$ state)\cite{Senthil_PRL_2003,Punk_PNAS_2015}, or various other topologically ordered versions of other familiar  phases.\cite{Sachdev_PRB_2016} 
More broadly, it has been suggested that a host of behaviors of highly correlated electronic systems can be best understood from the perspective of doped spin-liquids.\cite{Lee_RMP_2006,Fradkin_RMP_2015,Tocchio_PRB_2013,Gull_PRL_2013} 
However, a ``microscopic'' theory of QSLs is exceedingly difficult, as they seem to arise only in
 narrow portions of the generalized phase diagram where more typical broken symmetry states are suppressed by frustration, and in an  ``intermediate coupling'' regime  where neither the effective kinetic nor the interaction energy is dominant.

\begin{figure}
  \includegraphics[width=\linewidth]{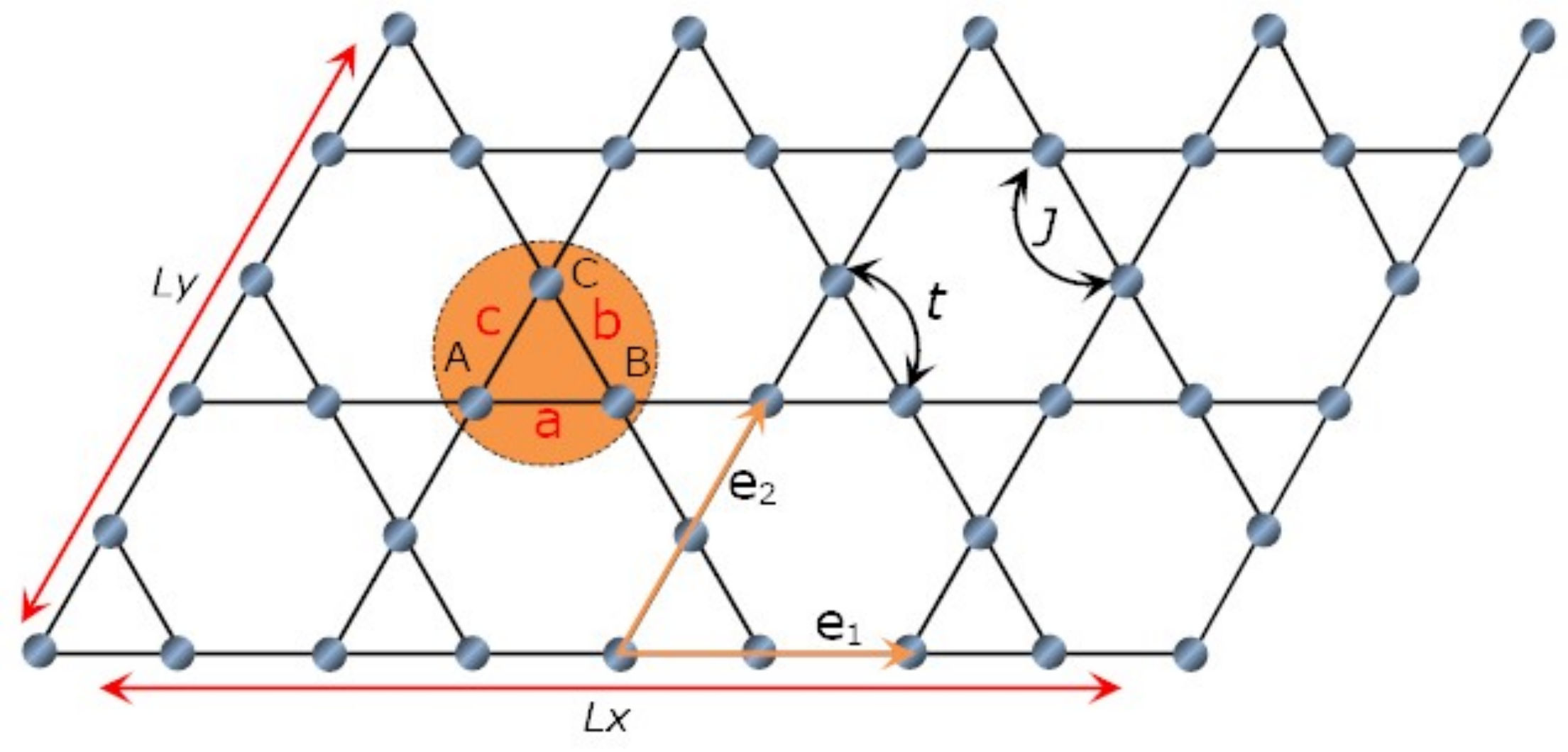}
  \caption{(Color online) The $t$-$J$ model on Kagome cylinder, where the electrons and spins are located at the vertices of the lattice (filled circles). $e_1$ and $e_2$ denote the two basis vectors of the lattice. Periodic (PBC) and open (OBC) boundary conditions are imposed along the $e_2$ and $e_1$ directions, respectively. Each unit cell (denoted by small triangle in the shaded region) has three sites (denoted by A, B and C) and three bonds (denoted by a, b and c).  $t$ and $J$ are hopping intergral and spin exchange interactions between nearest-neighbor (NN) sites. In referring to cylinders with this geometry, we refer to them as YC-$2L_y$, where $L_x$ and $L_y$ are the number of unit cells in the $e_1$ and $e_2$ directions, respectively.}\label{Fig:Kagome_Lattice}
\end{figure}

The spin 1/2 antiferromagnet on the Kagome lattice (depicted in Fig.\ref{Fig:Kagome_Lattice}) with nearest-neighbor (NN) Heisenberg interactions, i.e., $H_J$ in Eq.(\ref{Eq:ModelHamiltonian}), is 
geometrically frustrated. A number of numerical simulations  \cite{Jiang_PRL_2008,White_Science_2011,Jiang_NaturePhysics_2012,Depenbrock_PRL_2012,Gong_PRB_2015,Mei_arXiv1606_09639} have provided strong evidence that its ground state is a ``Z$_2$-QSL'' state with exponentially falling spin-spin correlations and a non-zero spin-gap, 
although some recent studies\cite{Ran_PRL_2007,Iqbal_PhysRevB.87.060405,Iqbal_PhysRevB.89.020407,Liao_arXiv1610_04727,Zaletel_arXiv_2016} have suggested that the true ground-state in the 2D limit may be a gapless (nodal) QSL.  The fact that the observed spin correlation lengths in the earlier references are short compared to the width of the ladders studied leaves little room to doubt that they reflect the properties of a spin-gapped state.  None-the-less, it is plausible that there are at least two distinct QSL phases - one gapped and another ungapped - that are very close in energy such that the balance between them can shift as a function of ladder width, geometry, or slight changes in parameters;  if this is the case, it could reconcile the two sets of findings while leaving open the issue of which QSL is the ground-state in the 2D limit.    

Independent of which QSL has the lowest energy in  2D, in the present study,  the fact that the spin correlation lengths we observe are several times shorter than the width of our cyllinders leaves little doubt that we are studying the properties of a doped, fully gapped $Z_2$ spin liquid. Experimentally, the celebrated material Herbertsmithite [$\rm ZnCu_3(OH)_6Cl_2$] is a realization of the two-dimensional Kagome antiferromagnet\cite{Hering_FRG_2016,Jiang_KAFM_Neutron_2016} where the copper ions carry spin S =1/ 2 magnetic moments which condense to form a QSL groundstate. Specifically, experimental evidence of  fractional spin excitations has been found in neutron scattering and strong indications of  
 a spin-gap are seen in NMR studies of single crystals. \cite{Han_Nature_2012,Fu_Science_2015}

The elementary excitations of a Z$_2$-QSL can be constructed\cite{Rokhsar_PRL_1988,  Senthil_JPA_2001} as combinations of a fermionic charge 0 spin 1/2 ``spinon,'' a charge e spin 0 bosonic holon, and a neutral topological ``vison.''  The statistics of these particles is a matter of convenience - for instance a fermionic holon can be constructed\cite{SAK_PhysRevB.39.259} as a boundstate of a bosonic holon and a vison, while a normal spin 1/2, charge e hole can be constructed as a boundstate of a spinon and a holon. (We consider only hole-doping of the QSL, so we can safely ignore negatively charged excitations.)  

If we assume that the states achieved by adding a net positive charge density $\delta\ll 1$ per site to a QSL insulator can be described in terms of  dilute excitations on top of a persistent background QSL, then a variety of possible ground-state phases are natural to consider:  If the lowest energy charged excitations are ordinary holes, then these can either remain itinerant, forming a Fermi liquid, FL$^\ast$,\ (which can easily be seen to be a distinct phase from  a usual Fermi liquid in that the Fermi surface only encloses an area corresponding to the density of doped holes, and so violates Luttinger's theorem) or if they crystalize,
they might  form an insulating  hole Wigner Crystal (hWC) with one doped hole per emergent unit cell. \footnote{Note that the hWC has a spin-1/2 per unit cell, and so would be expected to magnetically order unless the lattice structure of the hWC is such that this higher-order lattice supports a QSL groundstate.}  If, on the other hand, the lowest energy charged excitations are holons, then they could condense to form a conventional superconducting state with small superfluid density $\sim \delta$, or they could crystalize.  We refer to the later state as WC$^\ast$, which is distinguishable from hWC in that there are no low energy spin degrees of freedom.  One could also imagine that the lowest energy charged excitations are holon pairs (or equivalently, spin-singlet hole pairs), which if they crystalize would form an insulating WC of Cooper pairs.  Still more complicated phases, including supersolid phases and conducting charge density waves, could occur if a fraction of the charged excitations crystalize while others remain itinerant, or by condensing fermionic holons or bosonic spin 1/2 charge e holes.

\begin{figure}
  \includegraphics[width=\linewidth]{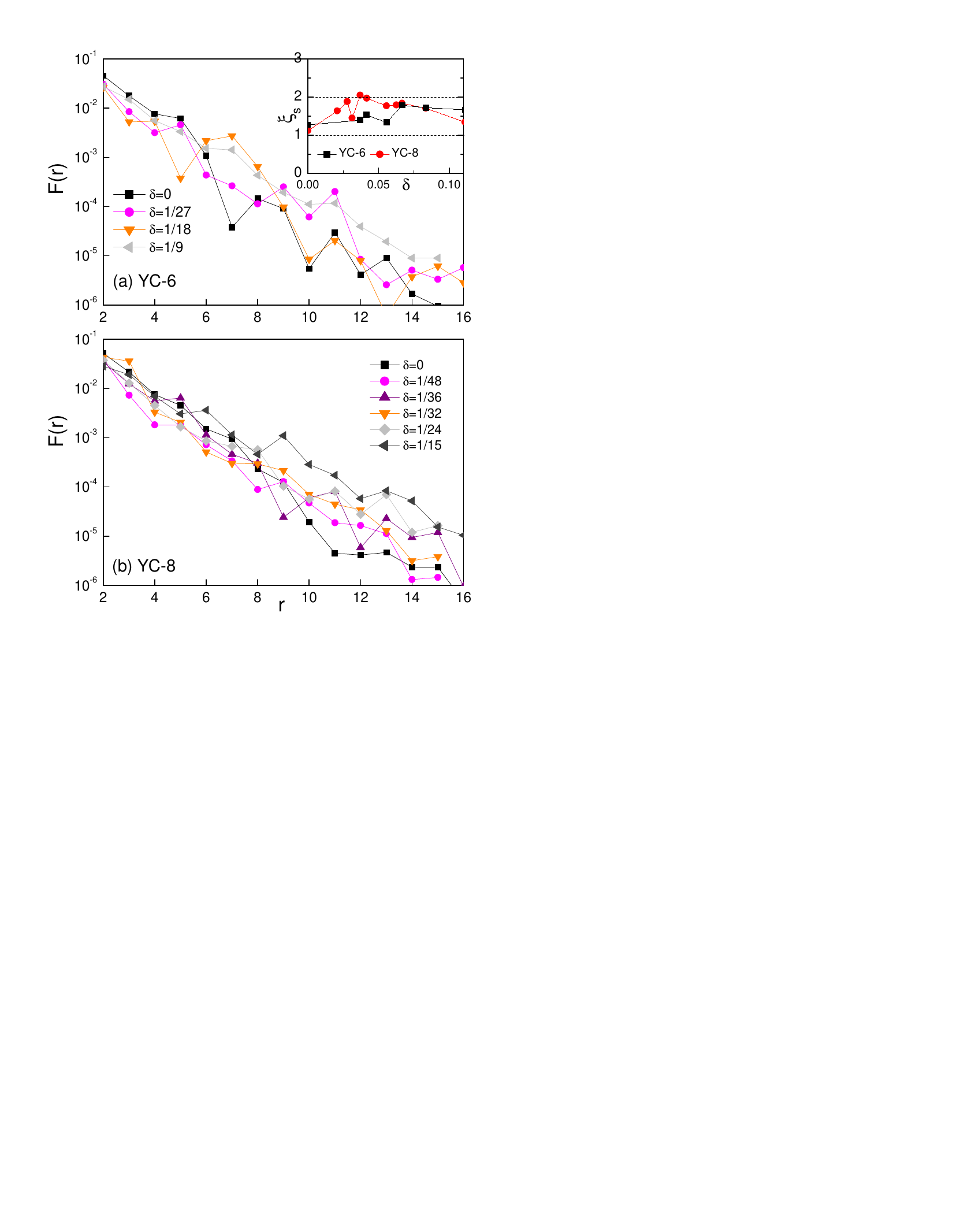}
  \caption{(Color online) The spin-spin correlation functions $\rm F(r)$ along the $\bold{e}_1$ direction for both (a) $\rm YC$-6 and (b) $\rm YC$-8 cylinders at different hole doping concentrations $\delta$. $r$ is the distance between two sites where the reference A site is in the middle of the system. The inset in (b) is the corresponding spin-spin correlation length $\xi_s$ as a function of $\delta$ for both $\rm YC$-6 and $\rm YC$-8 cylinders, by fitting $\rm F(r)$ to an exponential function $\rm F(r) \sim e^{-r/\xi_s}$.  Results plotted here are for $L_x$=16$\sim$24.}
  \label{Fig:Kagome_SpinCor}
\end{figure}

 \textbf{Principal Results:}  We find that the spin-spin correlation function (Fig. 2) is remarkably insensitive to doping.  Indeed, the spin correlation length, $\xi_s\lesssim 2$ lattice constants, is small compared to the circumference of the cylinders studied and to the mean separation between doped  holes.   The fact that they look little different than those for $\delta=0$ is consistent with viewing the system as a lightly doped QSL.   The expectation value of the charge density (Figs. 4 and 5) is inhomogeneous, and the amplitude of the charge density variations is relatively insensitive to system size, implying that the ground-state in the thermodynamic limit spontaneously breaks translation symmetry.  Moreover, for the most part, the charge density appears to favor a triangular lattice with one doped hole per unit cell.  Given the fact that there is no significant doping-induced magnetic order, we may identify this state as a WC$^\ast$.  However, the precise crystal structure of the WC$^\ast$ in the thermodynamic limit is not something we can infer with confidence, as in some cases, depending on the value of $\delta$ and the circumference of the cylinder,  we find a stripe crystal rather than a triangular lattice.  Given the strong evidence of crystalization, it is perhaps not surprising that we find that all superconducting correlations are extremely short-ranged (Fig. 3), with a correlation length $\xi_{SC} \lesssim 1.3$.
 \footnote{It is worth noting that the phases of a putative doped QSL was considered in another recent DMRG study of the $t$-$J$ model on the square lattice with a ratio of second and first neighbor exchange interactions, $J_2/J_1 \approx 1/2$ and with $\delta =1/8$.  While the systems considered there were much narrower than in the present study, clear evidence was found of a tendency to form a WC of singlet hole (or holon) pairs.\cite{Dodaro_arXiv_2016}}
 
\textbf{Model Hamiltonian:} %
We employ the density-matrix renormalization group (DMRG) \cite{White_PRL_1992} to investigate the ground state properties of the hole-doped Kagome antiferromagnet. The simplest model which captures the strong correlation physics on this lattice is the $t$-$J$ model (depicted in Fig.\ref{Fig:Kagome_Lattice}) which is defined by the Hamiltonian
\begin{eqnarray}\label{Eq:ModelHamiltonian}
H = -t\sum_{\langle ij\rangle \sigma}(c_{i\sigma}^\dagger c_{j\sigma} + h.c.) + J \sum_{\langle ij\rangle } \left (\mathbf{S}_i\cdot\mathbf{S}_j -\frac{1}{4}n_i n_j \right)
\end{eqnarray}
where $c_{i\sigma}^+$ ($c_{i\sigma}$) is the electron creation (annihilation) operator with spin-$\sigma$ on site $i$. $\vec{S}_i$ is the spin operator and $n_i=\sum_\sigma c_{i\sigma}^+ c_{i\sigma}$ is the electron number operator. $\langle ij\rangle$ denotes NN sites and the Hilbert space is constrained by the no-double occupancy condition, $n_i\leq 1$. At half-filling, i.e., $n_i=1$, the $t$-$J$ model reduces to the spin-$\frac{1}{2}$ antiferromagnetic Heisenberg model on the Kagome lattice.

The lattice geometry used in the DMRG simulations is depicted in Fig.\ref{Fig:Kagome_Lattice}, where $\bold{e}_1$ and $\bold{e}_2$ denotes the two basis vectors of the Kagome lattice. We consider Kagome cylinders with periodic (open) boundary condition in the $\bold{e}_2$ ($\bold{e}_1$)-direction. A cylinder geometry introduced Ref.\cite{Kolley_PRB_2015} (which we will refer to as YC)  is used such that 
one of the three bond orientations is along the $\bold{e}_2$-axis. Here, we focus on cylinders with width $L_y$ and length $L_x$, where $L_y$ and $L_x$ are the numbers of unit cells (and $2L_x$ and $2L_y$ are the number of sites) along the $\bold{e}_2$ and $\bold{e}_1$ directions, respectively. Notice that the unit cells at the right boundary of the cylinder contain only two sites (A and C) in order to reduce the boundary effects due to sharp edges. Following Ref.\cite{White_Science_2011,Kolley_PRB_2015}, we refer to the cylinders as $\rm YC$-$2L_y$, whose total number of sites is $N=L_y(3L_x+2)=N_u+2L_y$, where $N_u$ denotes the number of sites inside intact unit cells. In this paper, we will focus primarily on $\rm YC$-6 and $\rm YC$-8 cylinders, i.e., $L_y$=3 and $4$, with $L_x$=12 to 24. We have also considered $\rm YC$-10 cylinders, i.e., $L_y$=5, and found similar results (see \ref{Fig:Kagome_NiContour}(e)). As usual, the doping level of the system away from half-filling is defined as $\delta = N_h/N_u$, where $N_h$ is the number of holes. Although $N_u\neq N$ so that the average value of $\delta$ differs slightly from $\tilde{\delta}=N_h/N$, deep in the bulk, i.e., relatively far from the open boundaries, $\tilde{\delta}=\delta$.

For the present study, we focus on the lightly doped case with doping level $0\leq \delta\leq 11\%$. We set $J$=1 as an energy unit and consider $t$=3. The results also hold for other $t$. We perform up to 50 sweeps and keep up to $m$=10000 states in the DMRG block with a typical truncation error $\epsilon\sim 10^{-6}$ for $\rm YC$-6 cylinders, $\epsilon\sim 10^{-5}$ for $\rm YC$-8 cylinders and $\epsilon\sim 5\times 10^{-5}$ for $\rm YC$-10 cylinders. This leads to excellent convergence for the results that we report here which have been extrapolated to $m=\infty$ limit (see Supplementary Information).

\textbf{Spin-Spin correlations:} To describe the magnetic properties of the ground state of the systems, we calculate the spin-spin correlation functions defined as ${\rm F(r)}=\frac{1}{L_y} \sum_{y=1}^{L_y} |\langle \bold{S}_0 \cdot \bold{S}_r\rangle|$. Here $\bold{S}_0$ denotes the spin operator on the reference $A$-site in the middle of the cluster, while $\bold{S}_r$ runs over both $A$ and $B$ sites along the $\bold{e}_1$ -direction with the distance $r$ between them. At half-filling, i.e., $\delta=0$, the ground state of the system has been shown to be a QSL state \cite{Jiang_PRL_2008,White_Science_2011,Jiang_NaturePhysics_2012,Depenbrock_PRL_2012,Kolley_PRB_2015} with short-range spin-spin correlations. This is also confirmed by our present study where the spin-spin correlations $\rm F(r)$ for $\rm YC$-6 in Fig. \ref{Fig:Kagome_SpinCor}(a) and $\rm YC$-8 cylinders in Fig. \ref{Fig:Kagome_SpinCor}(b) both decay rapidly with distance $r$, and can be well fitted by an exponential function ${\rm F(r)} \sim e^{-r/\xi_s}$ with short correlation lengths $\xi_s$=1.1$\sim$1.3 lattice spacings.

\begin{figure}
  \includegraphics[width=\linewidth]{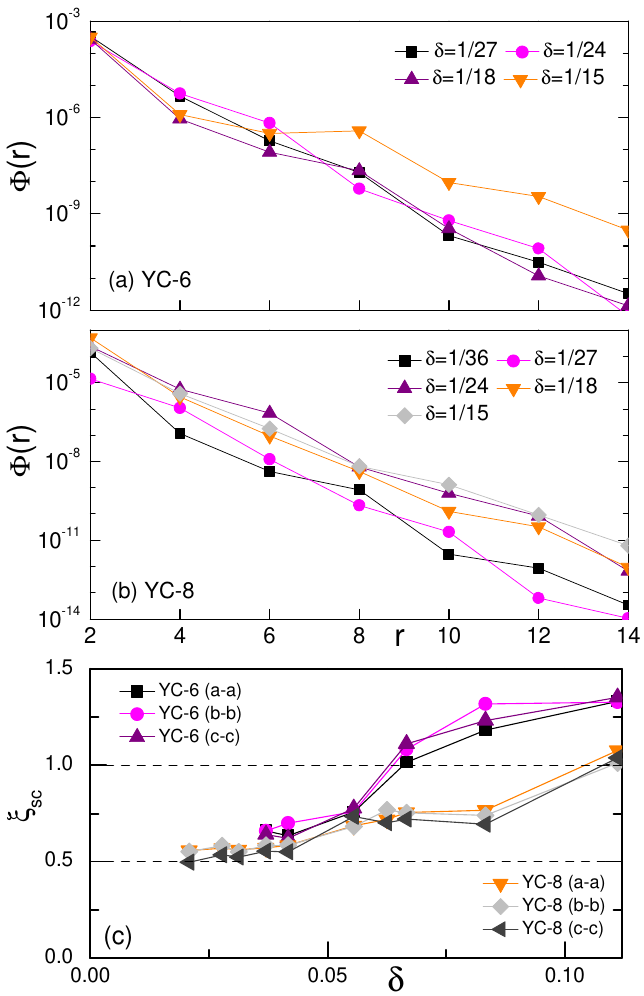}
  \caption{(Color online) The superconducting pair-field correlation functions $\rm \Phi_{aa}(r)$ along the $\bold{e}_1$ direction between ($a$-$a$)-bond for both (a) $\rm YC$-6 and (b) $\rm YC$-8 cylinders at different hole doping concentrations $\delta$. $r$ is the distance between two bonds with the reference bond in the middle of the system. Panel (c) shows the corresponding superconducting correlation length $\xi_{sc}$ as a function of $\delta$ for both $\rm YC$-6 and $\rm YC$-8 cylinders, by fitting $\rm \Phi(r)$ to an exponential function $\rm \Phi(r) \sim e^{-r/\xi_{sc}}$.  Here $L_x$=16$\sim$24.}\label{Fig:Kagome_SCCor}
\end{figure}

Upon doping, i.e., $\delta>0$, we find that the spin-spin correlations still remain short-ranged. The spin-spin correlations $\rm F(r)$ as a function of doping level $\delta$ at different system sizes are shown in Fig.(\ref{Fig:Kagome_SpinCor}). For all cases, we find that the spin-spin correlations $\rm F(r)$ decay exponentially with short correlation lengths $\xi_s$, although $\xi_s$ slightly depends on the doping level $\delta$ and lattice geometry. For both $\rm YC$-6 and $\rm YC$-8 cylinders, we find that $\xi_s$=1$\sim$2 lattice spacings as shown in the inset of Fig.(\ref{Fig:Kagome_SpinCor}) (a). Our results 
show that the ground states of the Kagome antiferromagnet upon doping remain spin singlet with short-range spin-spin correlations, similar to the QSL state at half-filling.

\textbf{Superconducting correlation:} %
 We have also investigated the possiblity of superconductivity. Since the ground state remains a spin-singlet state upon doping, we focus on spin-singlet superconductivity. The primary diagnostic of superconducting order is the pair-field correlator defined as
\begin{eqnarray}
\Phi_{\alpha \beta}(r)=\frac{1}{L_y}\sum_{y=1}^{L_y}|\langle \Delta^\dagger_\alpha(i_0) \Delta_\beta(i_0+r)\rangle|.\label{Eq:SC_Cor}
\end{eqnarray}
Here, $\Delta^\dagger_\alpha(i)$ is the spin-singlet pair-field creation operator given by
$\Delta_\alpha^\dagger(i) = \frac{1}{\sqrt{2}} \left( c^\dagger_{i \uparrow} c^\dagger_{i + \alpha \downarrow} - c^\dagger_{i \downarrow} c^\dagger_{i + \alpha \uparrow} \right)$,
where $\alpha$ denotes the bond type (see Fig.\ref{Fig:Kagome_Lattice}), i.e., a, b or c, with bond vectors defined as $\bold{a}=\bold{e}_1/2$, $\bold{c}=\bold{e}_2$ and $\bold{b}=(\bold{e}_2 - \bold{e}_1)/2$. $i_0$ is the index of the reference bond in the middle of the cluster, and $r$ is the distance between two bonds along the $\bold{e}_1$-direction.

In the present study, we find that $\Phi_{aa}$, $\Phi_{bb}$, $\Phi_{cc}$, $\Phi_{ab}$, $\Phi_{bc}$ and $\Phi_{ca}$ all decay exponentially for both $\rm YC$-6 and $\rm YC$-8 cylinders (see Fig.\ref{Fig:Kagome_SCCor}(a) and (b)).  In all cases we have studied, for large separations along the cylinder,   $1 \ll |r| \ll L_x$ where $ \bold{r} = r{\hat e}_1$, $\Phi$ can be well expressed as
$\Phi_{\alpha,\beta}(r) \sim e^{-|\bold{r}|/\xi^{\alpha\beta}_{sc}}$
from which we derive the values for superconducting correlation length $\xi^{\alpha\beta}_{sc}$. As shown in Fig.\ref{Fig:Kagome_SCCor}(c), $\xi_{sc}= 0.5\sim 1.3$ lattice spacings over all doping levels $0<\delta \leq 11\%$ we have explored. Therefore, our results suggest that there is no (quasi-) long-range superconductivity in doped QSL on the Kagome lattice.

\begin{figure}
  \includegraphics[width=\linewidth]{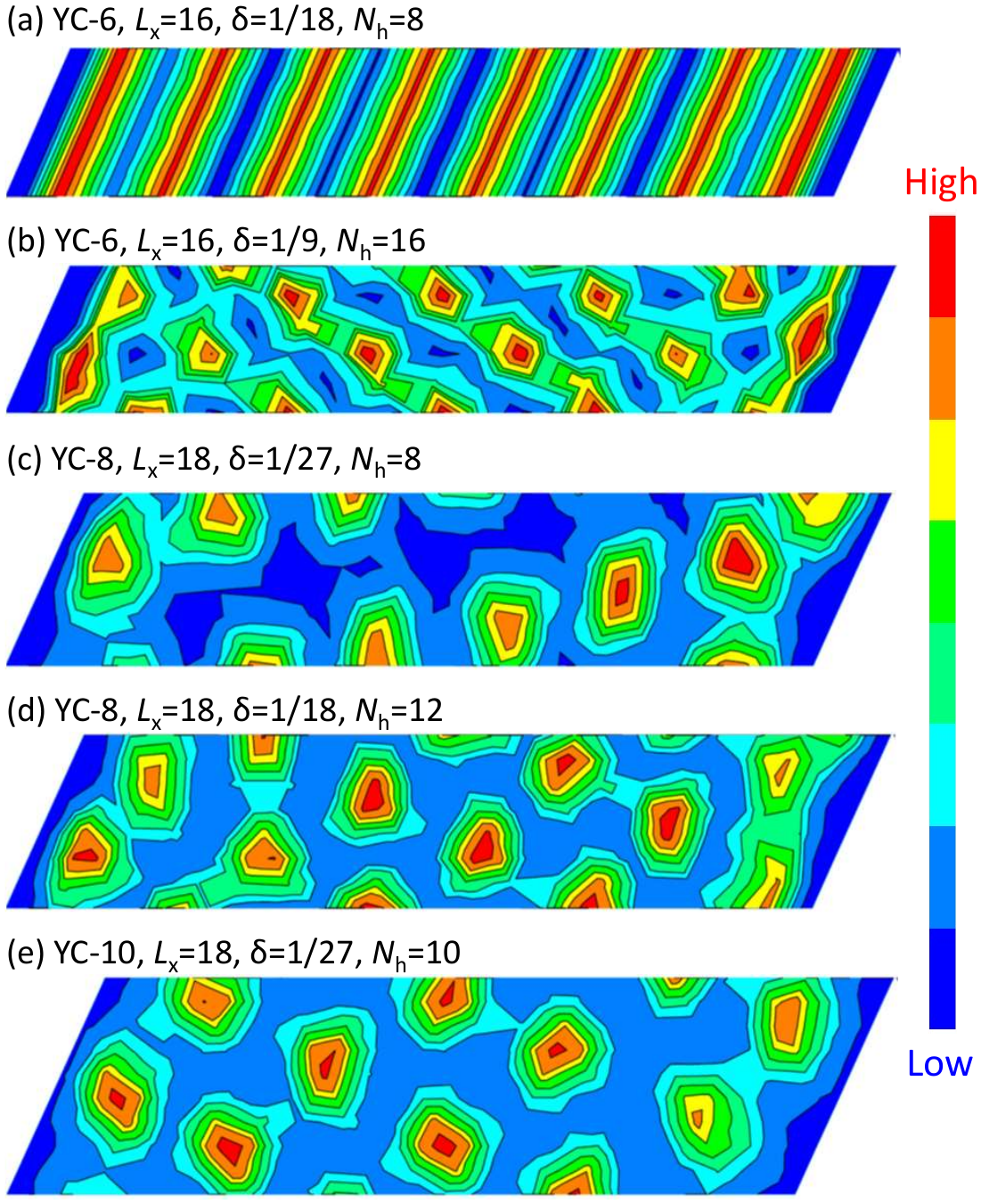}
  \caption{(Color online) The charge density profile $n_h(x,y)$ as a function of hole doping concentration $\delta$ for $\rm YC$-6 cylinders in (a)-(b), $\rm YC$-8 cylinders in (c)-(d) and $\rm YC$-10 cylinders in (e).}\label{Fig:Kagome_NiContour}
\end{figure}

\textbf{Charge density wave order:} %
Finally, we consider the charge density profile $n_h(x,y)=1-n(x,y)$, where $n(x,y)$ is the electron density on site $i=(x,y)$ with coordinations $x$ $\&$ $y$. Fig.\ref{Fig:Kagome_NiContour} shows some examples of $n_h(x,y)$ at different hole doping levels $\delta$ for both $\rm YC$-6 and $\rm YC$-8 cylinders. Clear CDW ordering is observed, although its pattern depends on both the lattice geometry (either $\rm YC$-6 or $\rm YC$-8) and doping level.
 There is unidirectional CDW order at low doping levels for $\rm YC$-6 cylinders, Fig.\ref{Fig:Kagome_NiContour}(a), but this appears to be special for $\rm YC$-6 geometry. For higher doping level for $\rm YC$-6 cylinders and all doping levels for $\rm YC$-8 cylinders, the CDW order resembles a two-dimensional Winger crystal, Fig.\ref{Fig:Kagome_NiContour}(b)-(d).

\begin{figure}
  \includegraphics[width=\linewidth]{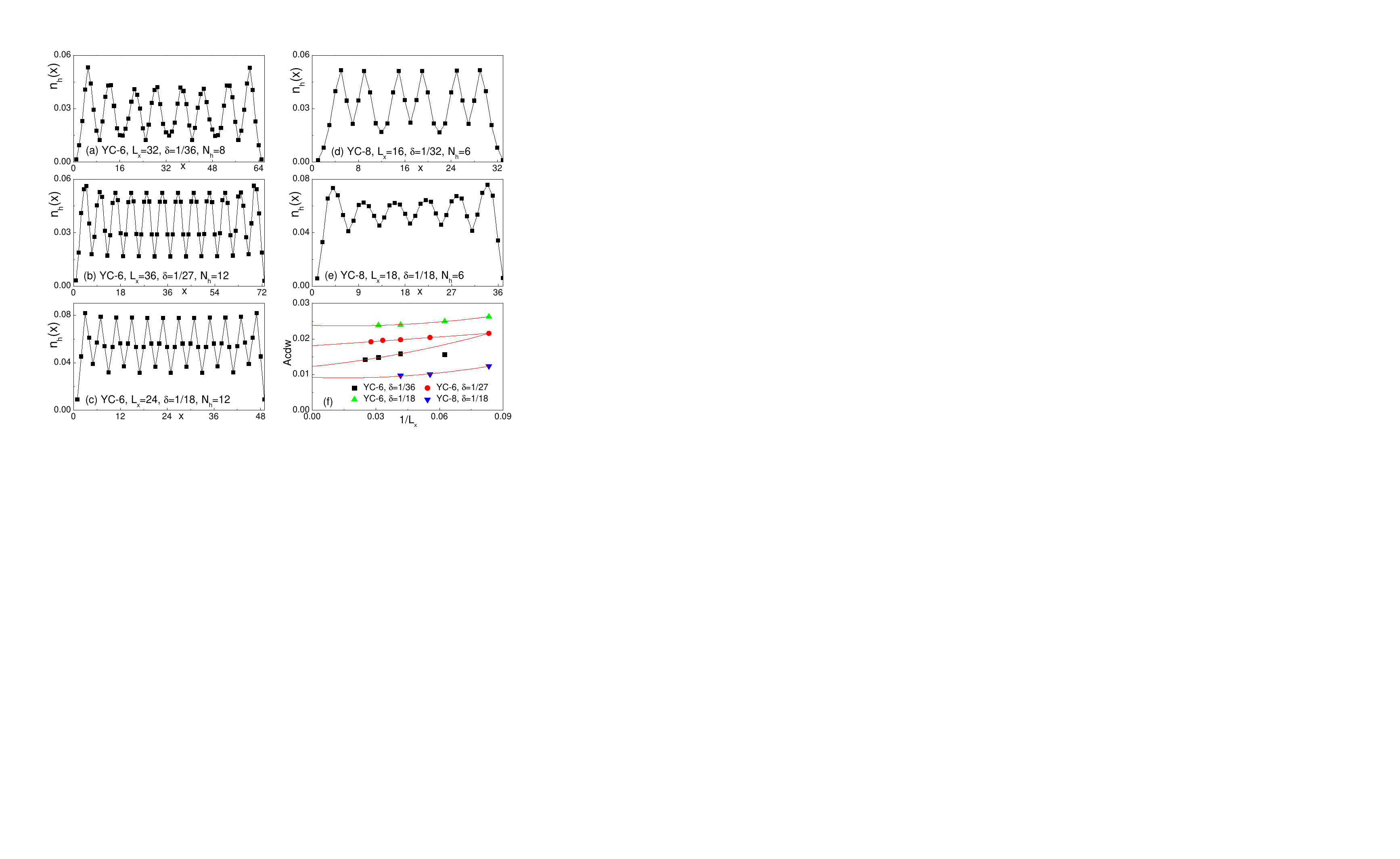}
  \caption{(Color online) The charge density profile $n_h(x)$ (includes both A and B sites) for the $t$-$J$ model  as a function of hole doping concentration $\delta$ for $\rm YC$-6 cylinders in (a)-(c) and $\rm YC$-8 cylinders in (d)-(e). (f) shows CDW order parameter $A_{cdw}$ by fitting the raw numberical data curve using function $n_h(x)=A_{cdw} \cos (Q_{cdw}+\theta)+\cdots$, where $A_{cdw}$ and $Q_{cdw}$ are CDW order parameter and ordering wave vector, respectively. A few points were removed from both ends in the fitting in order to minimize boundary effects.}\label{Fig:Kagome_Ni}
\end{figure}

Approximately, the doped system can be divided into new larger emergent unit cells, i.e. each containing one of the red stripes in Fig.\ref{Fig:Kagome_NiContour}(a) or one of the red-spots in Fig.\ref{Fig:Kagome_NiContour}(b)-(e); 
the number of emergent unit cells is equal to the number of doped holes at all doping levels. This is not a crystal  of Cooper pairs.

To determine whether the CDW order survives in the thermodynamic limit, we further calculate the averaged rung charge density defined by $n_h(x)=\frac{1}{L_y}\sum_{y=1}^{L_y}n_h(x,y)$. Examples of $n_h(x)$ at different $\delta$ for both $\rm YC$-6 and $\rm YC$-8 cylinders are plotted in Fig.\ref{Fig:Kagome_Ni}. Here, the existence of long-range CDW order in the ground state can be determiend by fitting the amplitude $A_{cdw}$ of the oscillation of the charge density $n_h(x)$ and extrapolating the value to $L_x=\infty$. To minimize the boundary effect, we have removed a few points from both ends in the fitting process. Examples of the extrapolation are given in Fig. \ref{Fig:Kagome_Ni} for system sizes $L_x=12 \sim 40$. The observed finite amplitude $A_{cdw}$ in the thermodynamic limit establishes the presence of long-range CDW order.

\textbf{Discussion:} %
In light of our observations, it is worth asking if there is an intuitive reason that the holons crystalize, rather than forming one of the possible quantum fluid states that have been proposed.  It is already clear from previous numerical studies that the QSL in the Kagome system is nearly degenerate with a number of possible valence-bond-crystalline phases.  It is thus natural to imagine that the holon is a highly structured particle, surrounded by a ``polaronic'' cloud of valence-bond-crystal like correlations.  

In Fig.S3 in the Supplemental Section, we show that two doped holes in cylinders of moderate length ($L_x$=12 \& 16) induce a strong and extended pattern of valence bond crystalline  order in their neighborhoods.  The most obvious consequence of this is that the holon effective mass is strongly renormalized (increased). Moreover, the induced valence bond order implies the existence of moderate range effective interactions between pairs of holons, which if they are repulsive can naturally lead to crystallization.  Note that in Fig.S2, we show that  the spin-gap in the presence of two doped holes (extrapolate to the $L_x\to \infty$ limit) is of the same order, although probably smaller  than in the undoped ladder;  this further corroborates our identification of this as a two holon state.

Since the Kagome antiferromagnet, i.e., $H_J$ in Eq.(\ref{Eq:ModelHamiltonian}), has been shown to be a realistic model to describe the celebrated Herbertsmithite \cite{Hering_FRG_2016,Jiang_KAFM_Neutron_2016}, our results may be directly relevant to the real material upon doping. Consistent with the present results, a recent experimental study has reported the absence of superconductivity in doped Kagome systems either with doped electrons \cite{Kelly_PRX_2016} or with doped holes \cite{Julien_PhysRevB.87.214423}. While we  have only studied finite cylinders, it is plausible that the results are representative of the thermodynamic limit, given the fact that the size of the cylinders, including both width $L_y$ and length $L_x$, are much larger than both the spin-spin and superconducting correlation lengths.

\textbf{Acknowledgement:}
We would like to thank Senthil Todadri, Young Lee, John Tranquada, Masaki Oshikawa, Arun Paramekanti, Roderich Moessner, Steve White, Zheng-Yu Weng, Hong Yao, Shenxiu Liu and especially Leon Balents for insightful discussions, and John Dodaro for making Fig.S3 (b) and (d). This work was supported by the Department of Energy, Office of Science, Basic Energy Sciences, Materials Sciences and Engineering Division, under Contract DE-AC02-76SF00515. Parts of the computing for this project was performed on the Sherlock cluster.


\begin{thebibliography}{36}
\expandafter\ifx\csname natexlab\endcsname\relax\def\natexlab#1{#1}\fi
\expandafter\ifx\csname bibnamefont\endcsname\relax
  \def\bibnamefont#1{#1}\fi
\expandafter\ifx\csname bibfnamefont\endcsname\relax
  \def\bibfnamefont#1{#1}\fi
\expandafter\ifx\csname citenamefont\endcsname\relax
  \def\citenamefont#1{#1}\fi
\expandafter\ifx\csname url\endcsname\relax
  \def\url#1{\texttt{#1}}\fi
\expandafter\ifx\csname urlprefix\endcsname\relax\def\urlprefix{URL }\fi
\providecommand{\bibinfo}[2]{#2}
\providecommand{\eprint}[2][]{\url{#2}}

\bibitem[{\citenamefont{Anderson}(1987)}]{Anderson_RVB_1987}
\bibinfo{author}{\bibfnamefont{P.~W.} \bibnamefont{Anderson}},
  \bibinfo{journal}{Science} \textbf{\bibinfo{volume}{235}},
  \bibinfo{pages}{1196} (\bibinfo{year}{1987}).

\bibitem[{\citenamefont{Kivelson et~al.}(1987)\citenamefont{Kivelson, Rokhsar,
  and Sethna}}]{Kivelson_PRB_1987}
\bibinfo{author}{\bibfnamefont{S.~A.} \bibnamefont{Kivelson}},
  \bibinfo{author}{\bibfnamefont{D.~S.} \bibnamefont{Rokhsar}},
  \bibnamefont{and} \bibinfo{author}{\bibfnamefont{J.~P.}
  \bibnamefont{Sethna}}, \bibinfo{journal}{Phys. Rev. B}
  \textbf{\bibinfo{volume}{35}}, \bibinfo{pages}{8865} (\bibinfo{year}{1987}).

\bibitem[{\citenamefont{Rokhsar and Kivelson}(1988)}]{Rokhsar_PRL_1988}
\bibinfo{author}{\bibfnamefont{D.~S.} \bibnamefont{Rokhsar}} \bibnamefont{and}
  \bibinfo{author}{\bibfnamefont{S.~A.} \bibnamefont{Kivelson}},
  \bibinfo{journal}{Phys. Rev. Lett.} \textbf{\bibinfo{volume}{61}},
  \bibinfo{pages}{2376} (\bibinfo{year}{1988}).

\bibitem[{\citenamefont{Laughlin}(1988)}]{Laughlin_PRL_1988}
\bibinfo{author}{\bibfnamefont{R.~B.} \bibnamefont{Laughlin}},
  \bibinfo{journal}{Phys. Rev. Lett.} \textbf{\bibinfo{volume}{60}},
  \bibinfo{pages}{2677} (\bibinfo{year}{1988}).

\bibitem[{\citenamefont{Wen and Lee}(1996)}]{Wen_PRL_1996}
\bibinfo{author}{\bibfnamefont{X.-G.} \bibnamefont{Wen}} \bibnamefont{and}
  \bibinfo{author}{\bibfnamefont{P.~A.} \bibnamefont{Lee}},
  \bibinfo{journal}{Phys. Rev. Lett.} \textbf{\bibinfo{volume}{76}},
  \bibinfo{pages}{503} (\bibinfo{year}{1996}).

\bibitem[{\citenamefont{Balents}(2010)}]{Balents2010}
\bibinfo{author}{\bibfnamefont{L.}~\bibnamefont{Balents}},
  \bibinfo{journal}{Nature} \textbf{\bibinfo{volume}{464}},
  \bibinfo{pages}{199} (\bibinfo{year}{2010}).

\bibitem[{\citenamefont{Senthil and Lee}(2005)}]{Senthil_PRB_2005}
\bibinfo{author}{\bibfnamefont{T.}~\bibnamefont{Senthil}} \bibnamefont{and}
  \bibinfo{author}{\bibfnamefont{P.~A.} \bibnamefont{Lee}},
  \bibinfo{journal}{Phys. Rev. B} \textbf{\bibinfo{volume}{71}},
  \bibinfo{pages}{174515} (\bibinfo{year}{2005}).

\bibitem[{\citenamefont{Senthil et~al.}(2003)\citenamefont{Senthil, Sachdev,
  and Vojta}}]{Senthil_PRL_2003}
\bibinfo{author}{\bibfnamefont{T.}~\bibnamefont{Senthil}},
  \bibinfo{author}{\bibfnamefont{S.}~\bibnamefont{Sachdev}}, \bibnamefont{and}
  \bibinfo{author}{\bibfnamefont{M.}~\bibnamefont{Vojta}},
  \bibinfo{journal}{Phys. Rev. Lett.} \textbf{\bibinfo{volume}{90}},
  \bibinfo{pages}{216403} (\bibinfo{year}{2003}).

\bibitem[{\citenamefont{Punk et~al.}(2015)\citenamefont{Punk, Allais, and
  Sachdev}}]{Punk_PNAS_2015}
\bibinfo{author}{\bibfnamefont{M.}~\bibnamefont{Punk}},
  \bibinfo{author}{\bibfnamefont{A.}~\bibnamefont{Allais}}, \bibnamefont{and}
  \bibinfo{author}{\bibfnamefont{S.}~\bibnamefont{Sachdev}},
  \bibinfo{journal}{Proceedings of the National Academy of Sciences}
  \textbf{\bibinfo{volume}{112}}, \bibinfo{pages}{9552} (\bibinfo{year}{2015}).

\bibitem[{\citenamefont{Patel et~al.}(2016)\citenamefont{Patel, Chowdhury,
  Allais, and Sachdev}}]{Sachdev_PRB_2016}
\bibinfo{author}{\bibfnamefont{A.~A.} \bibnamefont{Patel}},
  \bibinfo{author}{\bibfnamefont{D.}~\bibnamefont{Chowdhury}},
  \bibinfo{author}{\bibfnamefont{A.}~\bibnamefont{Allais}}, \bibnamefont{and}
  \bibinfo{author}{\bibfnamefont{S.}~\bibnamefont{Sachdev}},
  \bibinfo{journal}{Phys. Rev. B} \textbf{\bibinfo{volume}{93}},
  \bibinfo{pages}{165139} (\bibinfo{year}{2016}).

\bibitem[{\citenamefont{Lee et~al.}(2006)\citenamefont{Lee, Nagaosa, and
  Wen}}]{Lee_RMP_2006}
\bibinfo{author}{\bibfnamefont{P.~A.} \bibnamefont{Lee}},
  \bibinfo{author}{\bibfnamefont{N.}~\bibnamefont{Nagaosa}}, \bibnamefont{and}
  \bibinfo{author}{\bibfnamefont{X.-G.} \bibnamefont{Wen}},
  \bibinfo{journal}{Rev. Mod. Phys.} \textbf{\bibinfo{volume}{78}},
  \bibinfo{pages}{17} (\bibinfo{year}{2006}).

\bibitem[{\citenamefont{Fradkin et~al.}(2015)\citenamefont{Fradkin, Kivelson,
  and Tranquada}}]{Fradkin_RMP_2015}
\bibinfo{author}{\bibfnamefont{E.}~\bibnamefont{Fradkin}},
  \bibinfo{author}{\bibfnamefont{S.~A.} \bibnamefont{Kivelson}},
  \bibnamefont{and} \bibinfo{author}{\bibfnamefont{J.~M.}
  \bibnamefont{Tranquada}}, \bibinfo{journal}{Rev. Mod. Phys.}
  \textbf{\bibinfo{volume}{87}}, \bibinfo{pages}{457} (\bibinfo{year}{2015}).

\bibitem[{\citenamefont{Tocchio et~al.}(2013)\citenamefont{Tocchio, Lee,
  Jeschke, Valent\'{\i}, and Gros}}]{Tocchio_PRB_2013}
\bibinfo{author}{\bibfnamefont{L.~F.} \bibnamefont{Tocchio}},
  \bibinfo{author}{\bibfnamefont{H.}~\bibnamefont{Lee}},
  \bibinfo{author}{\bibfnamefont{H.~O.} \bibnamefont{Jeschke}},
  \bibinfo{author}{\bibfnamefont{R.}~\bibnamefont{Valent\'{\i}}},
  \bibnamefont{and} \bibinfo{author}{\bibfnamefont{C.}~\bibnamefont{Gros}},
  \bibinfo{journal}{Phys. Rev. B} \textbf{\bibinfo{volume}{87}},
  \bibinfo{pages}{045111} (\bibinfo{year}{2013}).

\bibitem[{\citenamefont{Gull et~al.}(2013)\citenamefont{Gull, Parcollet, and
  Millis}}]{Gull_PRL_2013}
\bibinfo{author}{\bibfnamefont{E.}~\bibnamefont{Gull}},
  \bibinfo{author}{\bibfnamefont{O.}~\bibnamefont{Parcollet}},
  \bibnamefont{and} \bibinfo{author}{\bibfnamefont{A.~J.}
  \bibnamefont{Millis}}, \bibinfo{journal}{Phys. Rev. Lett.}
  \textbf{\bibinfo{volume}{110}}, \bibinfo{pages}{216405}
  (\bibinfo{year}{2013}).

\bibitem[{\citenamefont{Jiang et~al.}(2008)\citenamefont{Jiang, Weng, and
  Sheng}}]{Jiang_PRL_2008}
\bibinfo{author}{\bibfnamefont{H.~C.} \bibnamefont{Jiang}},
  \bibinfo{author}{\bibfnamefont{Z.~Y.} \bibnamefont{Weng}}, \bibnamefont{and}
  \bibinfo{author}{\bibfnamefont{D.~N.} \bibnamefont{Sheng}},
  \bibinfo{journal}{Phys. Rev. Lett.} \textbf{\bibinfo{volume}{101}},
  \bibinfo{pages}{117203} (\bibinfo{year}{2008}).

\bibitem[{\citenamefont{Yan et~al.}(2011)\citenamefont{Yan, Huse, and
  White}}]{White_Science_2011}
\bibinfo{author}{\bibfnamefont{S.}~\bibnamefont{Yan}},
  \bibinfo{author}{\bibfnamefont{D.}~\bibnamefont{Huse}}, \bibnamefont{and}
  \bibinfo{author}{\bibfnamefont{S.}~\bibnamefont{White}},
  \bibinfo{journal}{Science} \textbf{\bibinfo{volume}{332}},
  \bibinfo{pages}{1173} (\bibinfo{year}{2011}).

\bibitem[{\citenamefont{Jiang et~al.}(2012)\citenamefont{Jiang, Wang, and
  Balents}}]{Jiang_NaturePhysics_2012}
\bibinfo{author}{\bibfnamefont{H.~C.} \bibnamefont{Jiang}},
  \bibinfo{author}{\bibfnamefont{Z.}~\bibnamefont{Wang}}, \bibnamefont{and}
  \bibinfo{author}{\bibfnamefont{L.}~\bibnamefont{Balents}},
  \bibinfo{journal}{Nature Physics} \textbf{\bibinfo{volume}{8}},
  \bibinfo{pages}{902} (\bibinfo{year}{2012}).

\bibitem[{\citenamefont{Depenbrock et~al.}(2012)\citenamefont{Depenbrock,
  McCulloch, and Schollw\"ock}}]{Depenbrock_PRL_2012}
\bibinfo{author}{\bibfnamefont{S.}~\bibnamefont{Depenbrock}},
  \bibinfo{author}{\bibfnamefont{I.~P.} \bibnamefont{McCulloch}},
  \bibnamefont{and}
  \bibinfo{author}{\bibfnamefont{U.}~\bibnamefont{Schollw\"ock}},
  \bibinfo{journal}{Phys. Rev. Lett.} \textbf{\bibinfo{volume}{109}},
  \bibinfo{pages}{067201} (\bibinfo{year}{2012}).

\bibitem[{\citenamefont{Gong et~al.}(2015)\citenamefont{Gong, Zhu, Balents, and
  Sheng}}]{Gong_PRB_2015}
\bibinfo{author}{\bibfnamefont{S.-S.} \bibnamefont{Gong}},
  \bibinfo{author}{\bibfnamefont{W.}~\bibnamefont{Zhu}},
  \bibinfo{author}{\bibfnamefont{L.}~\bibnamefont{Balents}}, \bibnamefont{and}
  \bibinfo{author}{\bibfnamefont{D.~N.} \bibnamefont{Sheng}},
  \bibinfo{journal}{Phys. Rev. B} \textbf{\bibinfo{volume}{91}},
  \bibinfo{pages}{075112} (\bibinfo{year}{2015}).

\bibitem[{\citenamefont{{Mei} et~al.}(2016)\citenamefont{{Mei}, {Chen}, {He},
  and {Wen}}}]{Mei_arXiv1606_09639}
\bibinfo{author}{\bibfnamefont{J.-W.} \bibnamefont{{Mei}}},
  \bibinfo{author}{\bibfnamefont{J.-Y.} \bibnamefont{{Chen}}},
  \bibinfo{author}{\bibfnamefont{H.}~\bibnamefont{{He}}}, \bibnamefont{and}
  \bibinfo{author}{\bibfnamefont{X.-G.} \bibnamefont{{Wen}}},
  \bibinfo{journal}{ArXiv e-prints}  (\bibinfo{year}{2016}),
  \eprint{1606.09639}.

\bibitem[{\citenamefont{Ran et~al.}(2007)\citenamefont{Ran, Hermele, Lee, and
  Wen}}]{Ran_PRL_2007}
\bibinfo{author}{\bibfnamefont{Y.}~\bibnamefont{Ran}},
  \bibinfo{author}{\bibfnamefont{M.}~\bibnamefont{Hermele}},
  \bibinfo{author}{\bibfnamefont{P.~A.} \bibnamefont{Lee}}, \bibnamefont{and}
  \bibinfo{author}{\bibfnamefont{X.-G.} \bibnamefont{Wen}},
  \bibinfo{journal}{Phys. Rev. Lett.} \textbf{\bibinfo{volume}{98}},
  \bibinfo{pages}{117205} (\bibinfo{year}{2007}).

\bibitem[{\citenamefont{Iqbal et~al.}(2013)\citenamefont{Iqbal, Becca, Sorella,
  and Poilblanc}}]{Iqbal_PhysRevB.87.060405}
\bibinfo{author}{\bibfnamefont{Y.}~\bibnamefont{Iqbal}},
  \bibinfo{author}{\bibfnamefont{F.}~\bibnamefont{Becca}},
  \bibinfo{author}{\bibfnamefont{S.}~\bibnamefont{Sorella}}, \bibnamefont{and}
  \bibinfo{author}{\bibfnamefont{D.}~\bibnamefont{Poilblanc}},
  \bibinfo{journal}{Phys. Rev. B} \textbf{\bibinfo{volume}{87}},
  \bibinfo{pages}{060405} (\bibinfo{year}{2013}).

\bibitem[{\citenamefont{Iqbal et~al.}(2014)\citenamefont{Iqbal, Poilblanc, and
  Becca}}]{Iqbal_PhysRevB.89.020407}
\bibinfo{author}{\bibfnamefont{Y.}~\bibnamefont{Iqbal}},
  \bibinfo{author}{\bibfnamefont{D.}~\bibnamefont{Poilblanc}},
  \bibnamefont{and} \bibinfo{author}{\bibfnamefont{F.}~\bibnamefont{Becca}},
  \bibinfo{journal}{Phys. Rev. B} \textbf{\bibinfo{volume}{89}},
  \bibinfo{pages}{020407} (\bibinfo{year}{2014}).

\bibitem[{\citenamefont{{Liao} et~al.}(2016)\citenamefont{{Liao}, {Xie},
  {Chen}, {Liu}, {Xie}, {Huang}, {Normand}, and
  {Xiang}}}]{Liao_arXiv1610_04727}
\bibinfo{author}{\bibfnamefont{H.~J.} \bibnamefont{{Liao}}},
  \bibinfo{author}{\bibfnamefont{Z.~Y.} \bibnamefont{{Xie}}},
  \bibinfo{author}{\bibfnamefont{J.}~\bibnamefont{{Chen}}},
  \bibinfo{author}{\bibfnamefont{Z.~Y.} \bibnamefont{{Liu}}},
  \bibinfo{author}{\bibfnamefont{H.~D.} \bibnamefont{{Xie}}},
  \bibinfo{author}{\bibfnamefont{R.~Z.} \bibnamefont{{Huang}}},
  \bibinfo{author}{\bibfnamefont{B.}~\bibnamefont{{Normand}}},
  \bibnamefont{and} \bibinfo{author}{\bibfnamefont{T.}~\bibnamefont{{Xiang}}},
  \bibinfo{journal}{ArXiv e-prints}  (\bibinfo{year}{2016}),
  \eprint{1610.04727}.

\bibitem[{\citenamefont{{He} et~al.}(2016)\citenamefont{{He}, {Zaletel},
  {Oshikawa}, and {Pollmann}}}]{Zaletel_arXiv_2016}
\bibinfo{author}{\bibfnamefont{Y.-C.} \bibnamefont{{He}}},
  \bibinfo{author}{\bibfnamefont{M.~P.} \bibnamefont{{Zaletel}}},
  \bibinfo{author}{\bibfnamefont{M.}~\bibnamefont{{Oshikawa}}},
  \bibnamefont{and}
  \bibinfo{author}{\bibfnamefont{F.}~\bibnamefont{{Pollmann}}},
  \bibinfo{journal}{ArXiv e-prints}  (\bibinfo{year}{2016}),
  \eprint{1611.06238}.

\bibitem[{\citenamefont{Hering and Reuther}(2016)}]{Hering_FRG_2016}
\bibinfo{author}{\bibfnamefont{M.}~\bibnamefont{Hering}} \bibnamefont{and}
  \bibinfo{author}{\bibfnamefont{J.}~\bibnamefont{Reuther}},
  \bibinfo{journal}{arXiv:1610.09149}  (\bibinfo{year}{2016}).

\bibitem[{\citenamefont{Jiang et~al.}(2017)\citenamefont{Jiang, Wen, and
  Lee}}]{Jiang_KAFM_Neutron_2016}
\bibinfo{author}{\bibfnamefont{H.~C.} \bibnamefont{Jiang}},
  \bibinfo{author}{\bibfnamefont{J.}~\bibnamefont{Wen}}, \bibnamefont{and}
  \bibinfo{author}{\bibfnamefont{Y.~S.} \bibnamefont{Lee}},
  \bibinfo{journal}{In preparation}  (\bibinfo{year}{2017}).

\bibitem[{\citenamefont{Han et~al.}(2012)\citenamefont{Han, Helton, Chu,
  Nocera, Rodriguez-Rivera, Broholm, and Lee}}]{Han_Nature_2012}
\bibinfo{author}{\bibfnamefont{T.~H.} \bibnamefont{Han}},
  \bibinfo{author}{\bibfnamefont{J.~S.} \bibnamefont{Helton}},
  \bibinfo{author}{\bibfnamefont{S.}~\bibnamefont{Chu}},
  \bibinfo{author}{\bibfnamefont{D.~G.} \bibnamefont{Nocera}},
  \bibinfo{author}{\bibfnamefont{J.~A.} \bibnamefont{Rodriguez-Rivera}},
  \bibinfo{author}{\bibfnamefont{C.}~\bibnamefont{Broholm}}, \bibnamefont{and}
  \bibinfo{author}{\bibfnamefont{Y.~S.} \bibnamefont{Lee}},
  \bibinfo{journal}{Nature} \textbf{\bibinfo{volume}{132}},
  \bibinfo{pages}{5570} (\bibinfo{year}{2012}).

\bibitem[{\citenamefont{Fu et~al.}(2015)\citenamefont{Fu, Imai, Han, and
  Lee}}]{Fu_Science_2015}
\bibinfo{author}{\bibfnamefont{M.}~\bibnamefont{Fu}},
  \bibinfo{author}{\bibfnamefont{T.}~\bibnamefont{Imai}},
  \bibinfo{author}{\bibfnamefont{T.-H.} \bibnamefont{Han}}, \bibnamefont{and}
  \bibinfo{author}{\bibfnamefont{Y.~S.} \bibnamefont{Lee}},
  \bibinfo{journal}{Science} \textbf{\bibinfo{volume}{350}},
  \bibinfo{pages}{655} (\bibinfo{year}{2015}).

\bibitem[{\citenamefont{Senthil and Fisher}(2001)}]{Senthil_JPA_2001}
\bibinfo{author}{\bibfnamefont{T.}~\bibnamefont{Senthil}} \bibnamefont{and}
  \bibinfo{author}{\bibfnamefont{M.~P.~A.} \bibnamefont{Fisher}},
  \bibinfo{journal}{Journal of Physics A: Mathematical and General}
  \textbf{\bibinfo{volume}{34}}, \bibinfo{pages}{L119} (\bibinfo{year}{2001}).

\bibitem[{\citenamefont{Kivelson}(1989)}]{SAK_PhysRevB.39.259}
\bibinfo{author}{\bibfnamefont{S.}~\bibnamefont{Kivelson}},
  \bibinfo{journal}{Phys. Rev. B} \textbf{\bibinfo{volume}{39}},
  \bibinfo{pages}{259} (\bibinfo{year}{1989}).

\bibitem[{\citenamefont{White}(1992)}]{White_PRL_1992}
\bibinfo{author}{\bibfnamefont{S.~R.} \bibnamefont{White}},
  \bibinfo{journal}{Phys. Rev. Lett.} \textbf{\bibinfo{volume}{69}},
  \bibinfo{pages}{2863} (\bibinfo{year}{1992}).

\bibitem[{\citenamefont{Kolley et~al.}(2015)\citenamefont{Kolley, Depenbrock,
  McCulloch, Schollw\"ock, and Alba}}]{Kolley_PRB_2015}
\bibinfo{author}{\bibfnamefont{F.}~\bibnamefont{Kolley}},
  \bibinfo{author}{\bibfnamefont{S.}~\bibnamefont{Depenbrock}},
  \bibinfo{author}{\bibfnamefont{I.~P.} \bibnamefont{McCulloch}},
  \bibinfo{author}{\bibfnamefont{U.}~\bibnamefont{Schollw\"ock}},
  \bibnamefont{and} \bibinfo{author}{\bibfnamefont{V.}~\bibnamefont{Alba}},
  \bibinfo{journal}{Phys. Rev. B} \textbf{\bibinfo{volume}{91}},
  \bibinfo{pages}{104418} (\bibinfo{year}{2015}).

\bibitem[{\citenamefont{Kelly et~al.}(2016)\citenamefont{Kelly, Gallagher, and
  McQueen}}]{Kelly_PRX_2016}
\bibinfo{author}{\bibfnamefont{Z.~A.} \bibnamefont{Kelly}},
  \bibinfo{author}{\bibfnamefont{M.~J.} \bibnamefont{Gallagher}},
  \bibnamefont{and} \bibinfo{author}{\bibfnamefont{T.~M.}
  \bibnamefont{McQueen}}, \bibinfo{journal}{Phys. Rev. X}
  \textbf{\bibinfo{volume}{6}}, \bibinfo{pages}{041007} (\bibinfo{year}{2016}).

\bibitem[{\citenamefont{Julien et~al.}(2013)\citenamefont{Julien, Simonet,
  Canals, Ballou, Hassan, Affronte, Garlea, Darie, and
  Bordet}}]{Julien_PhysRevB.87.214423}
\bibinfo{author}{\bibfnamefont{M.-H.} \bibnamefont{Julien}},
  \bibinfo{author}{\bibfnamefont{V.}~\bibnamefont{Simonet}},
  \bibinfo{author}{\bibfnamefont{B.}~\bibnamefont{Canals}},
  \bibinfo{author}{\bibfnamefont{R.}~\bibnamefont{Ballou}},
  \bibinfo{author}{\bibfnamefont{A.~K.} \bibnamefont{Hassan}},
  \bibinfo{author}{\bibfnamefont{M.}~\bibnamefont{Affronte}},
  \bibinfo{author}{\bibfnamefont{V.~O.} \bibnamefont{Garlea}},
  \bibinfo{author}{\bibfnamefont{C.}~\bibnamefont{Darie}}, \bibnamefont{and}
  \bibinfo{author}{\bibfnamefont{P.}~\bibnamefont{Bordet}},
  \bibinfo{journal}{Phys. Rev. B} \textbf{\bibinfo{volume}{87}},
  \bibinfo{pages}{214423} (\bibinfo{year}{2013}).

\bibitem[{\citenamefont{{Dodaro} et~al.}(2016)\citenamefont{{Dodaro}, {Jiang},
  and {Kivelson}}}]{Dodaro_arXiv_2016}
\bibinfo{author}{\bibfnamefont{J.~F.} \bibnamefont{{Dodaro}}},
  \bibinfo{author}{\bibfnamefont{H.-C.} \bibnamefont{{Jiang}}},
  \bibnamefont{and} \bibinfo{author}{\bibfnamefont{S.~A.}
  \bibnamefont{{Kivelson}}}, \bibinfo{journal}{ArXiv e-prints}
  (\bibinfo{year}{2016}), \eprint{1610.04300}.

\end{thebibliography}

\appendix 

\begin{center}
\noindent {\large {\bf Supplementary Information}}
\end{center}

\renewcommand{\thefigure}{S\arabic{figure}}
\setcounter{figure}{0}
\renewcommand{\theequation}{S\arabic{equation}}
\setcounter{equation}{0}

\begin{figure}
  \includegraphics[width=\linewidth]{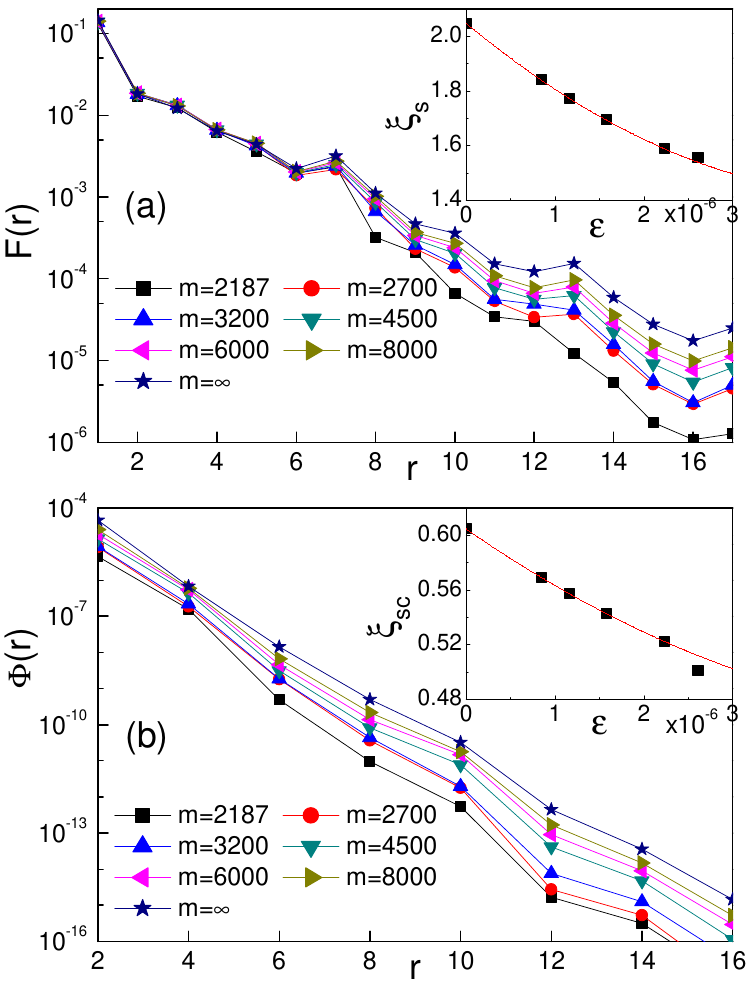}
  \caption{(Color online) (a) Spin-spin $\rm F(r)$ and (b) Superconducting correlation function $\rm \Phi(r)$ for $\rm YC$-8 cylinder at doping level $\delta=\frac{1}{27}$ keeping $m $= 2187 - 8000 states.  Insets show correlation length as a function of truncation error $\epsilon$. Here $L_x$=18.}\label{Fig:CorrelationScaling}
\end{figure}

\section{A: Extrapolated Correlation Funcation}\label{SI:Extrapolation}

In Fig. \ref{Fig:CorrelationScaling}(a), we show the spin-spin correlation function $\rm F(r)$ for $\rm YC$-8 cylinders with $L_x$=18 at $\delta=\frac{1}{27}$ hole doping concentration, keeping $m=2187 \sim 8000$ states. The navy blue line with star symbol shows the second-order polynomial extrapolation of the correlation function to $\epsilon \rightarrow 0$, i.e., $m=\infty$. The inset shows the correlation length $\xi_s$, fitted using the exponential function $\rm F(r) \sim e^{-r/\xi_s}$, as a function of truncation error $\epsilon$. The red line is a second-order fit of the data which gives the accurate $\xi_s$ in the $\epsilon \rightarrow 0$ limit.

In Fig. \ref{Fig:CorrelationScaling}(b), we show the superconducting correlation function for the same system. For simplicity, we define the averaged superconducting correlation function as $\rm \Phi(r) = [\Phi_{aa}(r)+\Phi_{bb}(r)+\Phi_{cc}(r)]/3$. The navy blue line with star symbols show the second-order polynomial extrapolation of $\rm \Phi(r)$ to $\epsilon \rightarrow 0$, and the inset shows the correlation length $\xi_{sc}$ as a function of $\epsilon$. The red line denotes the second-order fit.

\section{B: Spin Gap}\label{SI:SpinGap}

\begin{figure}
  \includegraphics[width=\linewidth]{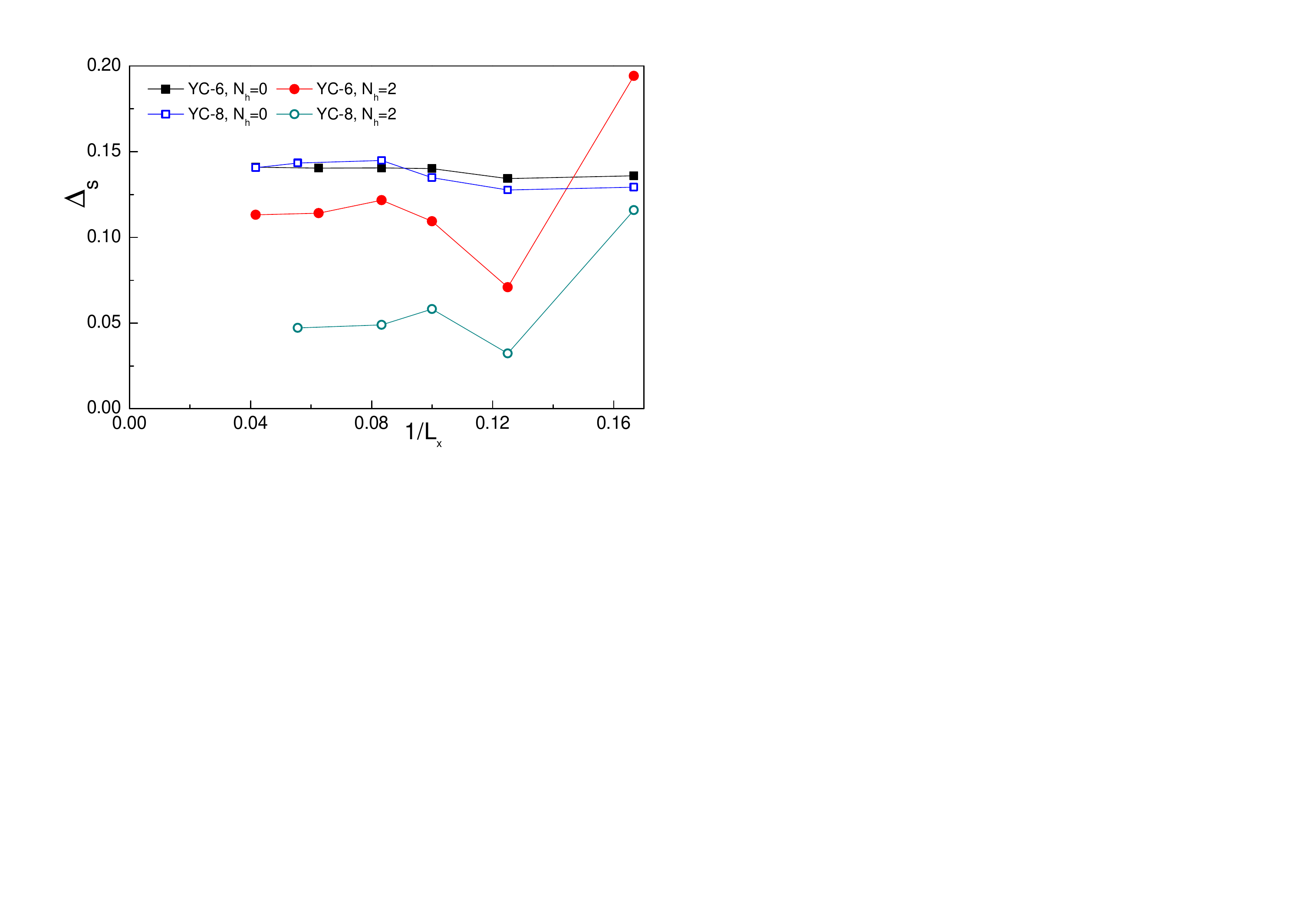}
  \caption{(Color online) Spin gap $\Delta_s$ for $\rm YC$-6 and $\rm YC$-8 cylinders at both half-filling ($N_h$=0) and doped with 2 holes ($N_h$=2).}\label{Fig:SpinGap}
\end{figure}

In Fig. \ref{Fig:SpinGap}, we shown the spin gap, defined as $\Delta_s(N_h)= E_0(N_h, S_z=1) - E_0(N_h, S_z=0)$ where $E_0(N_h, S_z)$ denotes the ground state energy for system with total spin $S_z$ doped with $N_h$ holes, for $\rm YC$-6 and $\rm YC$-8 cylinders at half-filling ($N_h$=0) and doped with 2 holes ($N_h$=2). At half-filling, $\Delta_s$ is finite for both cylinders, which is consistent with previous studies \cite{Jiang_PRL_2008,White_Science_2011,Jiang_NaturePhysics_2012,Depenbrock_PRL_2012}. When doped with 2 holes, $\Delta_s$ also appears to be finite for $\rm YC$-6 cylinder whose value is close to half-filling. However, for $\rm YC$-8 cylinder, we find that $\Delta_s$ appears to be much smaller than half-filling.

\begin{figure}
  \includegraphics[width=\linewidth]{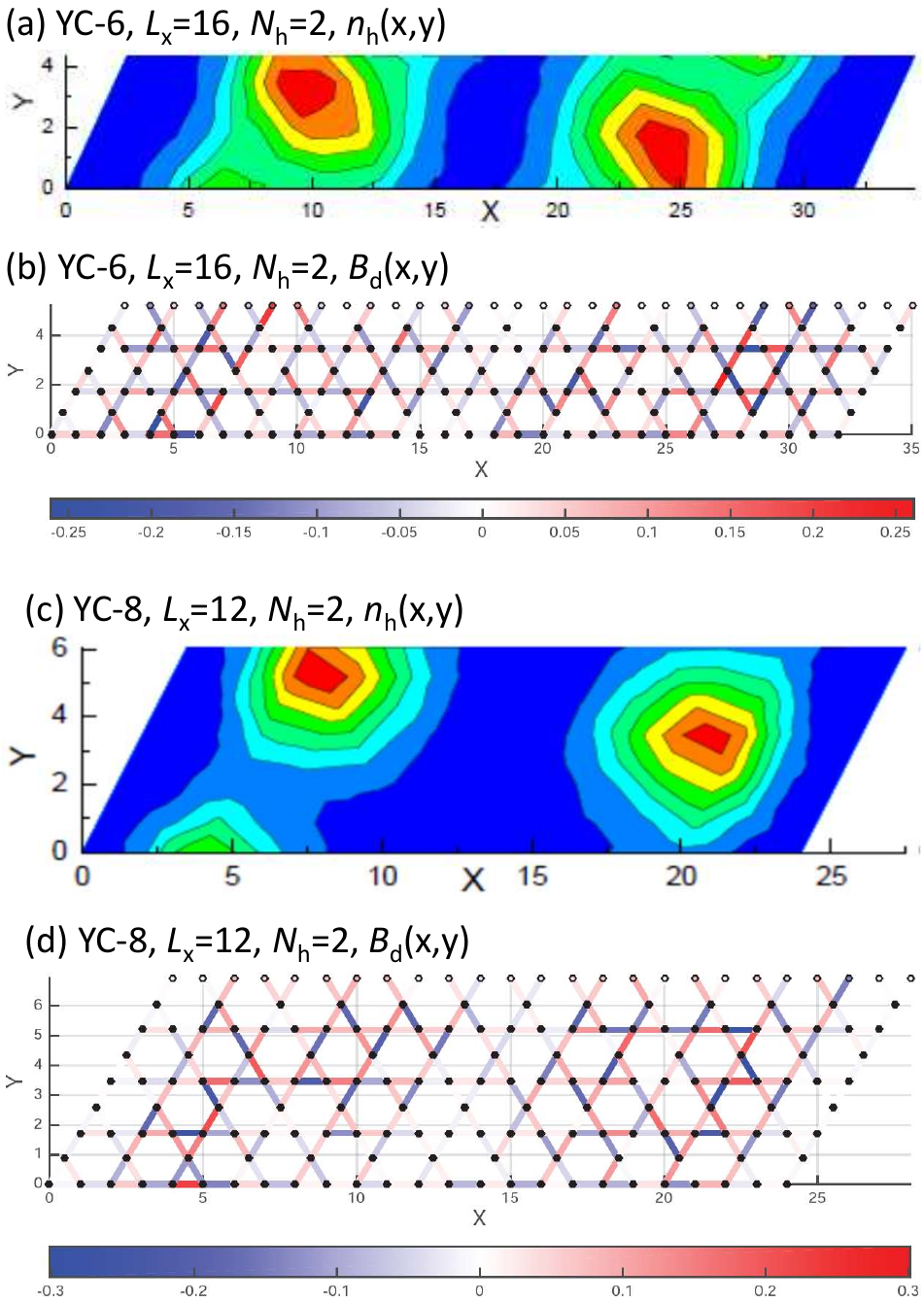}
  \caption{(Color online) The hole density profile $n_h(x,y)$ for (a) $\rm YC$-6 with $L_x$=16 and (c) $\rm YC$-8 with $L_x$=12 with $N_h$=2 doped holes. $B_d(x,y)$ is the difference in the valence bond strength between systems with $N_h$=2 doped holes and half-filling ($N_h$=0) for (b) $\rm YC$-6 with $L_x$=16 and (d) $\rm YC$-8 with $L_x$=12 cylinders.}\label{Fig:DimerDiff}
\end{figure}

\section{C: Valence Bond Order}\label{SI:VBO}%
In order to study the correlations that are associated with the  dressed holon, we have studied the properties of two doped holes in a system of a moderate size  so that the charge density associated with the two holons are  well localized near opposite ends of the cylinder, far enough from each other so that they hardly overlap. To look for local valence-bond crystalline correlations, we measure the difference in the valence bond strength  defined as $B_d(i,j)=\langle \vec{S}_i\cdot \vec{S}_j \rangle_{N_h=2}$ - $\langle \vec{S}_i\cdot \vec{S}_j \rangle_{N_h=0}$, where $i$ and $j$ are nearest neighbor sites and the subscripts indicate that the expectation values are taken in the ground-states with $N_h$=2  and $N_h=0$ doped holes, respectively. 
We show both the expectation value of the charge density (Figs. \ref{Fig:DimerDiff} (a) \& (c)) and $B_d(x,y)$  (Figs. \ref{Fig:DimerDiff} (b) \& (d)) for a $\rm YC$-6 system with $L_x$=16 and a $\rm YC$-8 system with $L_x$=12. Clearly, the holon is an extended object with a diameter of roughly 5 lattice constants in which    strong local valence bond crystalline correlations are manifest.

\end{document}